\documentstyle[epsfig]{aipproc}
\newbox\grsign \setbox\grsign=\hbox{$>$} 
\newdimen\grdimen \grdimen=\ht\grsign
\newbox\laxbox \newbox\gaxbox
\setbox\gaxbox=\hbox{\raise.5ex\hbox{$>$}\llap
     {\lower.5ex\hbox{$\sim$}}}\ht1=\grdimen\dp1=0pt
\setbox\laxbox=\hbox{\raise.5ex\hbox{$<$}\llap
     {\lower.5ex\hbox{$\sim$}}}\ht2=\grdimen\dp2=0pt
\def\gax{\mathrel{\copy\gaxbox}}

\newcommand{\epsB}{\mbox{$\epsilon_B$}}

\begin{document}
\title{Optical/Multiwavelength Observations of GRB Afterglows}

\author{Titus J. Galama$^*$} \address{$^*$Astronomy, MS 105-24,
California Institute of Technology, Pasadena, CA 91125\thanks{The
author is supported by the Sherman Fairchild Foundation.}}

\maketitle

\begin{abstract}
I review $\gamma$-ray burst optical/multiwavelength afterglow
observations since 1997, when the first counterparts to GRBs were
discovered.  I discuss what we have learned from multiwavelength
observations of GRB afterglows in relation to the `standard' fireball
plus relativistic blast-wave models. To first order the `standard'
model describes the afterglow observations well, but a wealth of
information can be gathered from the deviations of GRB afterglow
observations from this `standard' model. These deviations provide
information on the nature of the progenitor and on the physics of GRB
production. In particular I focus on the possible connection of GRBs
to supernovae, on jet and circumstellar wind models, on the early-time
afterglow, and on the emission from the reverse shock.
\end{abstract}

\section{Introduction}
%
%
%
%
\par\noindent Fireball-plus-relativistic blast-wave models
predict low-energy radiation following GRBs (see, e.g,
\cite{mr97}). This radiation has been dubbed the `afterglow'.
The basic model is that of a point explosion: a large amount of
energy, $\sim 10^{52-53}$ ergs is released in a compact region (less
than a light millisecond across), which leads to a `fireball', an
optically thick radiation-electron-positron plasma with initial energy
much larger than its rest mass that expands ultra-relativistically
(see, e.g., \cite{pir99} for an extensive review). The GRB may be due
to a series of `internal shocks' that develop in the relativistic
ejecta before they collide with the ambient medium. When the fireball
runs into the surrounding medium a `forward shock' ploughs into the
medium and heats it, and a `reverse shock' does the same to the
ejecta. As the forward shock is decelerated by increasing amounts of
swept-up material it produces a slowly fading `afterglow' of X rays,
followed by ultraviolet, optical, infrared, millimetre, and radio
radiation. As the reverse shock travels through the ejecta it may give
rise to a bright optical flash.

Models for the origin of GRBs that (in principle) can provide the
required energies, are the neutron star-neutron star (e.g.,
\cite{eic+89}) and neutron star-black hole mergers
\cite{moc+93,ls74,npp92}, white dwarf collapse \cite{uso92}, and core
collapses of very massive stars (`failed' supernovae or hypernovae
\cite{woo93,pac98}).

This review consists of two parts. In the first part, I discuss
several confirmations of the relativistic nature of GRB events and
discuss the generally good agreement between the `standard' fireball
plus relativistic blast wave model and the observations of GRB
afterglows. In the second part I then proceed to discuss the `devious'
deviations of some GRB afterglows from this standard model, and
discuss the wealth of information that we can gather from them. In
particular I discuss what such deviations may tell us about the nature
of the progenitor and about the physics of GRB production.

\section{Confirmation of the relativistic blast-wave model}

\subsection{The forward shock \label{sec:for}}

Let us first concentrate on the forward shock and assume slow cooling
(the bulk of the electrons do not radiate a significant fraction of
their own energy and the evolution is adiabatic); this appears
applicable to some observed GRB afterglows at late times ($t >$ 1 hr).

The electrons are assumed to be accelerated, in the forward shock, to
a power-law distribution of electron Lorentz factors, $N(\gamma_{\rm
e}) \propto \gamma_{\rm e}^{-p}$, with some minimum Lorentz factor
$\gamma_{\rm m}$. Then, the synchrotron spectrum of such a
distribution of electrons is a power law with $F_{\nu} \propto
\nu^{1/3}$ up to a maximum, $F_{\rm m}$, at the peak frequency
$\nu_{\rm m}$ (corresponding to the minimum Lorentz factor
$\gamma_{\rm m}$). Above $\nu_{\rm m}$ it is a power law, $F_{\nu}
\propto \nu^{-(p-1)/2}$, up to the cooling frequency, $\nu_{\rm
c}$. Electrons with energies $\gamma_{\rm e}m_{\rm e}c^2 > \gamma_{\rm
c}m_{\rm e}c^2$, where $\gamma_{\rm c}$ is the electron Lorentz factor
associated with the cooling frequency $\nu_{\rm c}$, radiate a
significant fraction of their energy and thereby cause a
spectral transition; above $\nu_{\rm c}$ we have $F_{\nu} \propto
\nu^{-p/2}$.  Synchrotron self absorption causes a steep cutoff of the
spectrum at low frequencies, $\nu < \nu_{\rm a}$ ($F_{\nu} \propto
\nu^2$ if $\nu_{\rm a} < \nu_{\rm m}$), where $\nu_{\rm a}$ is the
synchrotron self absorption frequency. Thus, the spectrum consists of
four distinct power-law regimes, seperated by three break frequencies:
(i) the self absorption frequency, $\nu_{\rm a}$, (ii) the peak
frequency, $\nu_{\rm m}$, and (iii) the cooling frequency, $\nu_{\rm
c}$ (see Fig. \ref{fig:spec}).

The simplest assumption is that of spherical symmetry and a constant
ambient density.  For example, if GRBs are the result of the merger of
a compact binary system (such as a double neutron star or a neutron
star-black hole binary system), then we would expect the fireball to
encounter a homogeneous ambient medium.  In that case the afterglow
can be described by the spectral shape described above combined with
the following scalings $\nu_{\rm m} \propto t_{\rm obs}^{-3/2}$,
$\nu_{\rm c} \propto t_{\rm obs}^{-1/2}$, $\nu_{\rm a} \propto t_{\rm
obs}^0 = {\rm constant}$, $F_{\rm m} \propto t_{\rm obs}^0 = {\rm
constant}$ (see \cite{spn98} and \cite{wg99} for details).

\subsubsection{The first X-ray and optical counterparts}

\begin{figure}[h!] 
\centerline{\epsfig{file=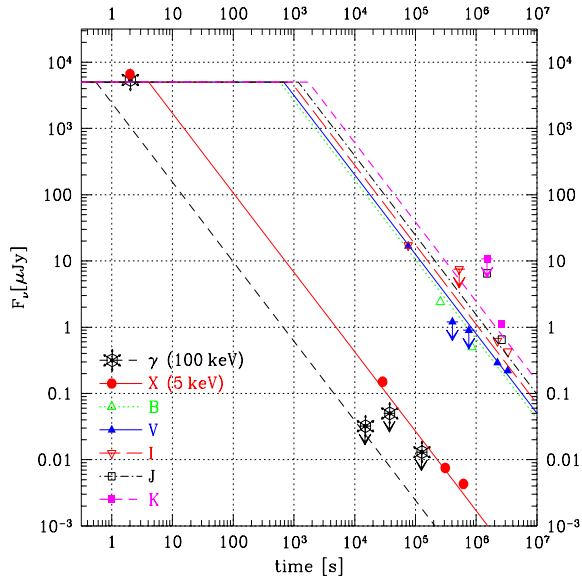,height=3.in,width=3.in}}
\vspace{10pt}
\caption{The light curves of GRB\,970228 from gamma rays to near
infrared (from [14]). To first order the light curves are
power laws and the offsets between them satisfy the expectations from
the model.}
\label{fig:wrm97}
\end{figure}

As both the afterglow's spectrum and the temporal evolution of the
break frequencies $\nu_{\rm a}$, $\nu_{\rm m}$, $\nu_{\rm c}$ are, in
this model, power laws, the evolution of the flux is also a power law
in time. For example, for $\nu_{\rm m} \leq \nu \leq \nu_{\rm c}$, the
decay of the flux is {\scriptsize $F_{\nu} \propto t_{\rm
obs}^{-3(p-1)/4}$}, and the power law spectral slope $\alpha$ relates
to the spectral slope $\beta$ as $\alpha = -3/2 \beta$. A stringent
test of the relativistic blast-wave model came with the discovery of
the first X-ray \cite{cfh+97} and optical \cite{pgg+97} counterparts
to GRB\,970228.  Several authors \cite{wrm97,rei97,wax97} showed that
to first order the model describes the X-ray and optical afterglow
very well (see Fig.  \ref{fig:wrm97}).

Detection of absorption features in the OT's spectrum of GRB\,970508
\cite{mdk+97} established that this event was at a redshift greater
than $z = 0.835$, showing that GRBs are located at cosmological
distances and are thus extremely powerful events.  This was also the first
GRB with a radio counterpart \cite{fkn+97}. The radio light curves
(8.5 and 4.9 GHz) show large variations on time scales of less than a
day, but these damp out after one month. This finds a viable
explanation in interstellar scintillation (irregular plasma refraction
by the interstellar medium between the source and the observer). The
damping of the fluctuations can then be understood as the effect of
source expansion on the diffractive interstellar scintillation. Thus a
source size of roughly 10$^{17}$ cm was derived (at 3 weeks),
corresponding to a mildly relativistic expansion of the shell
\cite{fkn+97}.

\begin{figure}[h!] 
\centerline{\epsfig{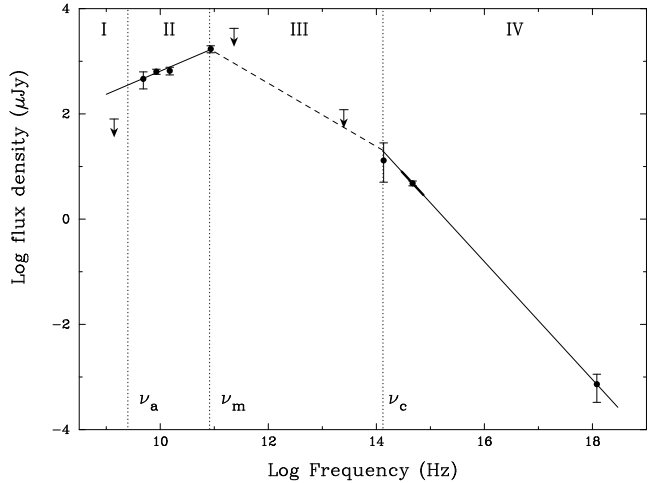}}
\vspace{10pt}
\caption{The X-ray to radio spectrum of GRB\,970508 on May 21.0 UT
(12.1 days after the event). The location of the break frequencies
$\nu_{\rm a}$, $\nu_{\rm m}$ and $\nu_{\rm c}$, inferred from
transitions in the light curves and from spectra of the afterglow, are
indicated (from [26]). }
\label{fig:spec}
\end{figure}

GRB\,970508 remains one of the best observed afterglows: the radio
afterglow was visible at least 368 days (and at 2.5 sigma on day 408.6
\cite{fwk99}), and the optical afterglow up to $\sim$ 450 days
(e.g. \cite{fpg+99,bdg+98,ggv+98a,cas+98}). In addition millimeter
\cite{bkg+98}, infrared and X-ray \cite{pir+98} counterparts were
detected, and it is the first GRB for which a spectral transition in
the optical/near IR range was found \cite{ggv+98a,gwb+98}; this
transition is interpreted as the effect of the passage of the cooling
frequency through the optical/near IR passbands. These multiwavelength
observations allowed the reconstruction of the broad radio to X-ray
spectrum for this GRB \cite{gwb+98} (see Fig. \ref{fig:spec}). It is
found that the `standard' model provides a successful and consistent
description of the afterglow observations over nine decades in
frequency, ranging in time from the event until several months later
\cite{gwb+98}. The synchrotron afterglow spectrum of this GRB allows
measurement of the electron energy spectrum $p$, the three break
frequencies ($\nu_{\rm a}$, $\nu_{\rm m}$ and $\nu_{\rm c}$), and the
flux at the peak, $F_{\rm m}$. For GRB\,970508 the redshift, $z$, is
also known, and all blast wave parameters could be deduced: the total
energy (per unit solid angle) E = 3.5$\times10^{52}$ erg, the ambient
(nucleon) density $n = 0.030$, the fraction of the energy in electrons
$\epsilon_{\rm e} = 0.12$ and that of the magnetic field \epsB = 0.089
\cite{wg99}. The numbers themselves are uncertain by an order of
magnitude (see e.g., \cite{gps+99}), but the result shows that the
`standard' model fits the expectations very well.

\subsection{The reverse shock}

\begin{figure}
\centerline{\epsfig{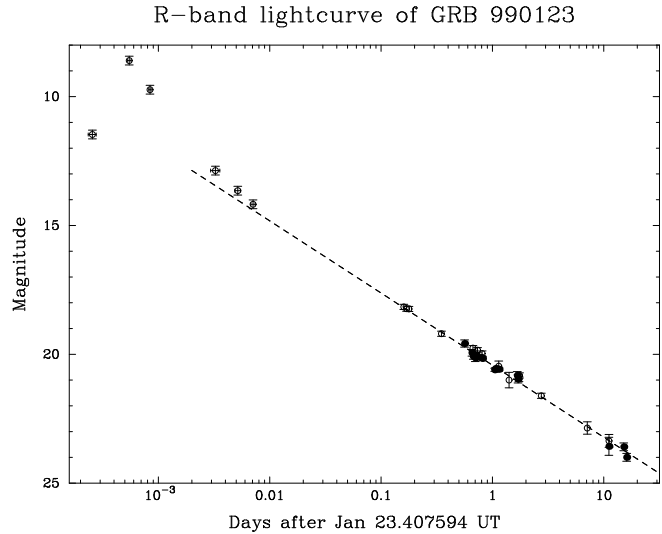}}
\vspace{10pt}
\caption{R-band light curve of the afterglow of GRB\,990123. The ROTSE
data show that the optical light curve peaked at m$_V \sim 9$
[28]. The dashed line indicates a power law fit to the light
curve (for $t > 0.1$ days), which has exponent $-1.12 \pm 0.03$ (from
[29]). \label{fig:0123}}
\end{figure}

The ROTSE telescope obtained its first images only 22 seconds after
the start of GRB\,990123 (i.e. during the GRB), following a
notification received from the BATSE aboard the Compton-satellite. The
ROTSE observations show that the optical light curve peaked at m$_V
\sim 9$ magnitudes some 60 seconds after the event
\cite{abb+99}. After maximum a fast decay follows for at least 15
minutes. The late-time afterglow observations show a more gradual
decline \cite{gbw+99,kul+99,cas+99,fat+99,sp99b} (see
Fig. \ref{fig:0123}).

The redshift $z = 1.6$, inferred from absorption features in the OT's
spectrum, implies that the optical flash would have been as bright as
the full moon had the GRB occured in the nearby galaxy M31
(Andromeda).  If one assumes that the emission detected by ROTSE comes
from a non-relativistic source of size $ct$, then the observed
brightness temperature $T_b\gax 10^{17}$\,K of the optical flash
exceeds the Compton limit of $10^{12}$\,K, confirming the highly
relativistic nature of the GRB source \cite{gbw+99}.

The ROTSE observations show that the prompt optical and $\gamma$-ray
light curves do not track each other \cite{abb+99}. In addition,
detailed comparison of the prompt optical emission with the BATSE
spectra of GRB\,990123 (at three epochs for which both optical and
gamma-ray information is available) shows that the ROTSE emission is
not a simple extrapolation of the GRB spectrum to much lower energies
\cite{gbw+99,bri+99}. 

Emission from the reverse shock is predicted to peak near the optical
waveband during or just after the GRB \cite{mr97,sp99a}. The observed
properties of GRB\,990123 appear to fit this model quite well
\cite{gbw+99,sp99b,mr99}. If this interpretation is correct,
GRB\,990123 would be the first burst in which all three emitting
regions have been seen: internal shocks causing the GRB, the reverse
shock causing the prompt optical flash, and the forward shock causing
the afterglow.  The emissions thus arise from three different emitting
regions, explaining the lack of correlation between the GRB, the
prompt optical and the late-time optical emission \cite{gbw+99} (but
see \cite{men+99}).

\section{Deviations \label{sec:grbsn}}

As discussed in the previous Section, the `standard' model explains
the multiwavelength observations of GRB afterglows very well. Now that
we have a basic understanding of GRB afterglows it is interesting to
consider what we can learn (and what we have learned in the past year)
from the observational departures from the `standard' model. 

\subsection{The GRB/Supernova connection \label{sec:grbsn}}

A direct consequence of the collapsar model is that GRBs are expected
to be accompanied by supernovae (SNe).

The first evidence for a possible GRB/SN connection was provided by
the discovery by Galama et al. \cite{gvv+98} of SN\,1998bw in the
error box of GRB\,980425. The temporal and spatial coincidence of
SN\,1998bw with GRB\,980425 suggest that the two phenomena are
related \cite{gvv+98,kul+98}.  GRB\,980425 is most certainly not a
typical GRB: the redshift of SN\,1998bw is 0.0085 and the
corresponding $\gamma$-ray peak luminosity of GRB\,980425 and its
total $\gamma$-ray energy budget are about a factor of $\sim$ 10$^5$
smaller than those of `normal' GRBs. Such SN-GRBs may well be the most
frequently occuring GRBs in the Universe.

\begin{figure}
\centerline{\epsfig{file=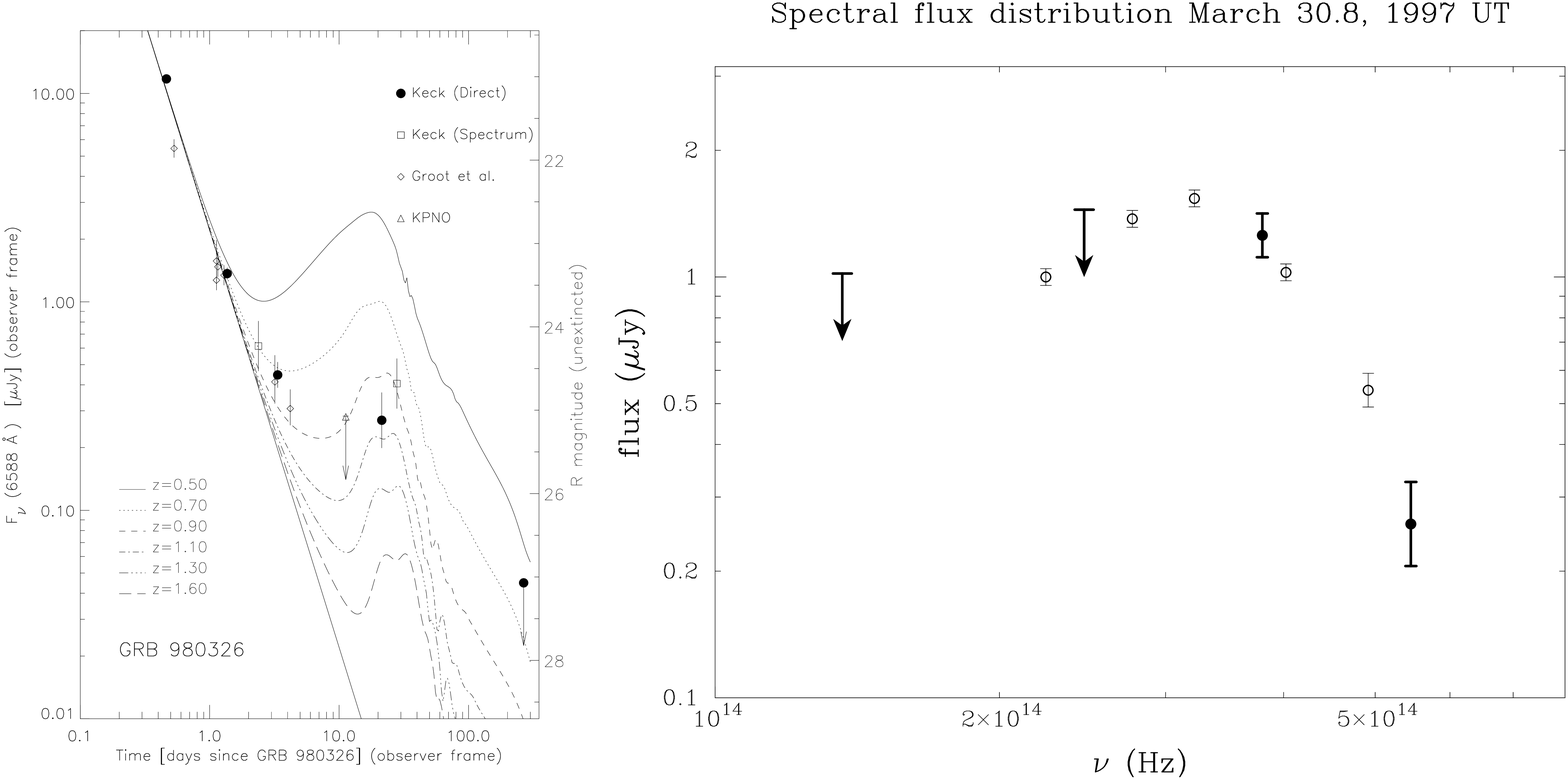,width=12.cm}}
\caption{Left: R-band light curve of GRB\,980326 and the sum of an
initial power-law decay plus Ic supernova light curve for redshifts
ranging from $z = 0.50$ to $z=1.60$ (from [40]). Right: The
broad-band spectrum of the OT of GRB\,970228 at March 30.8, 1997 UT
($\bullet$ and upper-limit arrow).  Also shown is the spectral flux
distribution of SN\,1998bw ($\circ$) redshifted to the redshift of
GRB\,970228 ($z = 0.695$). The similarity of the spectral flux
distributions is remarkable (from [43]). \label{fig:sn}}
\end{figure}

Bloom et al. \cite{bkd+99} realized that the late-time red spectrum
and the late-time rebrightnening of the light curve of GRB\,980326 are
possible evidence that at late times the emission is dominated by an
underlying supernova. The authors find that a template supernova light
curve, provided by the well-studied type I$_{b/c}$ SN\,1998bw provides
an adequate description of the observations (see Fig. \ref{fig:sn}).

In fact, the behavior of GRB\,970228 already showed first indications
that the standard model was not sufficient to describe the
observations in detail \cite{ggv+97}. The early-time decay of the
optical emission is faster than that at later times and, as the source
faded, it showed an unexpected reddening \cite{ggv+97}. Indeed, Galama
et al. \cite{ggv+97} conclude that although the initial behavior is in
agreement with the `standard' model, the subsequent behavior is harder
to explain. It was not until Bloom et al. \cite{bkd+99} discussed
evidence for a supernova-like emission acompanying GRB\,980326 that
the behavior of GRB\,970228 was better understood. Also for
GRB\,970228 the late-time light curve and reddening of the transient
can be well explained by an initial power-law decay modified at late
times by SN\,1998bw-like emission \cite{rei99,gtv+99} (see
Fig. \ref{fig:sn}).

The relation between distant GRBs like GRB\,980326 and GRB\,980425/SN
1998bw is unclear. Is SN\,1998bw a different phenomenon or a more
local and lower energy equivalent?  Are all afterglows consistent
with such a phenomenon?  The answer to the latter question requires
detailed analysis of existing data on GRB afterglows, but more
convincing evidence may be provided by future observations of GRB
afterglows around the time of the SN emission maximum.

 
\subsection{Collimated outflow (jet) and/or circumstellar wind model}

If, as suggested by the evidence for a GRB/SN connection (see
Sect. \ref{sec:grbsn}), at least some GRBs are produced by the core
collapse of massive stars to black holes, then the circumburst
environment will have been influenced by the strong wind of the
massive progenitor star. For a constant wind speed the circumstellar
density falls as $n\propto r^{-2}$, where $r$ is the radial
distance. In this, so called, circumstellar wind model, the afterglow
can be described by the same synchrotron spectral shape (see
Sect. \ref{sec:for}), but with different scalings for the break
frequencies and the peak flux: $\nu_{\rm m} \propto t_{\rm
obs}^{-3/2}$, $\nu_{\rm c} \propto t_{\rm obs}^{+1/2}$, $\nu_{\rm a}
\propto t_{\rm obs}^{-3/5}$, $F_{\rm m} \propto t_{\rm obs}^{-1/2}$
(see \cite{cl99a,cl99b} for details).

Due to relativistic beaming only a small portion of the emitting
surface with opening angle 1/$\gamma$ is visible. As the fireball
evolves $\gamma$ decreases and the beaming angle will eventually
exceed the angular size of the collimated outflow (the size of the
jet). In this jet model, we then expect to see an increase in the
decay rate. Slightly later the jet begins a lateral expansion, which
causes a further steepening of the light curve.  In this case the
scalings for the break frequencies are: $\nu_{\rm m} \propto t_{\rm
obs}^{-2}$, $\nu_{\rm c} \propto t_{\rm obs}^{0} = {\rm constant}$,
$\nu_{\rm a} \propto t_{\rm obs}^{-1/5}$, $F_{\rm m} \propto t_{\rm
obs}^{-1}$ (see for details \cite{rho99,sph99,pm99}). At late times,
when the evolution is dominated by the spreading of the jet the decay
is as fast as F$_{\nu}(t) \propto t^{-p} \sim t^{-2.2}$, where $p$ is
the power-law index of the electron energy spectrum.

{\bf Non `standard' behavior}: The optical and X-ray light curves of
GRB\,970508 show a maximum that is reached around 1 day and is
followed by characteristic power-law decaying light curves.  The onset
of the X-ray flare roughly coincides with that of the optical bump
\cite{pir+98}. This behavior is not yet well understood. Panaitescu et
al. \cite{pan+98} have tested several possible models to explain the
flare: (i) by continued energy injection from the central source, (ii)
by ejecta with a range of Lorentz factors. (iii) as the effect of a
jet that is observed slightly off-center, and (iv) by the encounter of
a shell of dense ambient material.

The afterglow peak flux F$_{\rm m}$ of GRB\,970508 decays with time;
it is $\sim$ 1700 $\mu$Jy at 86 GHz at $\sim$ 12 days, while only
$\sim$ 700 $\mu$Jy at 8.5 GHz at $\sim$ 60 days
\cite{fkn+97,bkg+98,gwb+98}. Also, the self-absorption frequency
$\nu_{\rm a}$ evolves to lower frequencies. However, in the `standard'
model the peak flux and the self-absorption frequency would remain
constant in time. Again, these features have several possible
explanations: (i) the effect of collimated outflow (ii) the effect of
a circumstellar wind, or (iii) the transition from an
ultra-relativistic to a non-relativistic evolution \cite{fwk99}. Note
however, that the `standard' model and the circumstellar wind model
predict a distinctively different evolution of the cooling break
$\nu_{\rm c}$; the observed evolution for GRB\,970508 fits the
`standard' model well and is hard to reconcile with the wind model.

{\bf Fast decaying afterglows}: GRB\,980326 was the first example of a
rapidly decaying afterglow \cite{ggv+98b}. Unfortunately, no attempt
was made to observe the X-ray afterglow, and the optical spectral
information is only sparse. It was not until GRB\,980519 that it was
decisively found that the rapidly decaying afterglow could not be
understood in the terms of the `standard' model; the relation between
the spectral slope and the temporal decay is not as expected from the
`standard' model. The observations can either be explained by a jet
\cite{sph99,hkp+99} or by a circumstellar wind model \cite{cl99a}.
Radio observations of GRB\,980519 are well described by a wind model,
but cannot decisively reject the jet model \cite{fks+99}. The reason
that it is hard to distinguish the different models is because of the
absense of high quality data; afterglows are faint. Future radio
observations at early and late times may allow to decisivly
distinguish the models.

Perhaps the actual light-curve transition (from a regular to a fast
decay caused by `seeing' the edge of the jet) has been observed in the
optical afterglow of GRB\,990123 \cite{kul+99,cas+99,fat+99}.
However, no evidence for such an increase of the decay rate was found
in near-infrared K-band observations \cite{kul+99}. A similar
transition was better sampled in afterglow data of GRB\,990510;
optical observations of GRB\,990510, show a clear steepening of the
rate of decay of the light between $\sim$ 3 hours and several days
\cite{hbf+99,sgk+99} to roughly F$_{\nu}(t) \sim t^{-2.2}$. Together
with radio observations, which also reveal a transition, it is found
that the transition is very much frequency-independent; this virtually
excludes explanations in terms of the passage of the cooling
frequency, but is what is expected in case of beaming
\cite{hbf+99}. Harrison et al.\cite{hbf+99} derive a jet opening angle
of $\theta = 0.08$, which for this burst would reduce the total energy
in $\gamma$ rays to $\sim 10^{51}$ erg.

\subsection{The early afterglow and the reverse shock}

The radio observations of GRB\,990123 show a brief flare at one day
after the event \cite{gbw+99,kfs+99}.  Such radio behavior is unique,
both for its early appearance as well as its rapid decline.  The flare
has been suggested to be due to the reverse shock
\cite{sp99b,fks+99}. However, understanding the full evolution still
requires interpretation in terms of the forward shock and a jet in
addition to the reverse shock.  An alternative interpretation in terms
of emission by the forward shock only is also consistent with the
observations \cite{gbw+99}. This interpretation is also not without
problems; the spectrum is required to be relatively flat around the
maximum. In this interpretation the energy density of the magnetic
field is very low \epsB $< 10^{-6}$, similar to what is derived for
GRB\,980703 \cite{vgo+99}. The differences in afterglow behavior may
thus reflect variations in the magnetic-field strength in the forward
shock \cite{gbw+99}. Other possibilities have been put forward: an
explanation in terms of the forward shock and a jet \cite{wdl99} and
an explanation in terms of the forward shock and a dense ambient
medium \cite{dl99}.  Interestingly, observations of the light curve at
times between 15 min. and several hours could distinguish between some
of the models; this is the region of transition from early times,
where the emission is believed to be due to the reverse shock, to late
times where the emission of the forward shock is dominant. The
imminent launch of HETE-2 will provide the unique possibility to study
this time window, by providing accurate localizations to the community
within minutes after the events.

\section{Conclusions}
Although the `standard' model describes the afterglow observations
well, a wealth of information is provided by the deviations of GRB
afterglows from the `standard' model; in particular, by
the possible connection of GRBs to supernovae, by possible evidence
for collimated outflow and circumstellar winds, by the early-time
afterglow and by the emission from the reverse shock.

\bibliographystyle{cite}

\end{document}